\UseRawInputEncoding
\documentclass[prd,twocolumn,showpacs,floatfix,amsmath,nofootinbib,amssymb,floatfix]{revtex4}
\usepackage{graphicx,color,dcolumn,booktabs,bm,multirow}
\usepackage{longtable,lscape}
\usepackage{txfonts}
 \usepackage{enumitem}
\usepackage{overpic}
\usepackage{amssymb}
\usepackage{indentfirst}
\usepackage{feynmf}   
\usepackage{slashed}  
\usepackage{cases}
\usepackage{color}
\usepackage{multirow}
\usepackage{epstopdf}
\usepackage{graphicx,color,dcolumn,booktabs,bm}

\usepackage[colorlinks,
            citecolor=blue,
            anchorcolor=red,
            menucolor=red,
            linkcolor=red,
            filecolor=red,
            runcolor=red,
            urlcolor=blue,
            frenchlinks=red]{hyperref}

\begin{document}

\title{Production of $Z_{cs}$ in $B$ and $B_s$ decay}
\author{Qi Wu$^{1}$}\email{wu\_qi@pku.edu.cn}
\author{Dian-Yong Chen$^{2,3}$ \footnote{Corresponding author}}\email{chendy@seu.edu.cn}
\author{Wen-Hua Qin$^{4}$}\email{qwh@qfnu.edu.cn}
\author{Gang Li$^{4}$}\email{gli@qfnu.edu.cn}
\affiliation{$^1$School of Physics and Center of High Energy Physics, Peking University, Beijing 100871, China\\
$^2$School of Physics, Southeast University, Nanjing 210094, China\\
$^3$Lanzhou Center for Theoretical Physics, Lanzhou University, Lanzhou 730000, China\\
$^4$College of Physics and Engineering, Qufu Normal University, Qufu 273165, China}

\begin{abstract}
In the present work, we investigate the production of $Z_{cs}^+$ in $B^+$ and $B_s^0$ decay, where $Z_{cs}^+$ is assigned as a $D_s^{+} \bar{D}^{\ast0} + D_s^{\ast +}\bar{D}^0$ molecular state. By using an effective Lagrangian approach, we evaluate the branching ratio of $B^0_s\rightarrow K^- Z^+_{cs}$ and $B^+\rightarrow \phi Z^{+}_{cs}$ via the triangle loop mechanism. The estimated branching fractions of $B^0_s\rightarrow K^- Z^+_{cs}$ and $B^+\rightarrow \phi Z^{+}_{cs}$ are an order of $10^{-4} $ and $10^{-5}$, respectively. The ratio of these two branching fraction is estimated to be about 5, which indicate that the $B_s^0 \to K^\pm  Z^\mp_{cs} \to K^+ K^- J/\psi$ may be a better process of searching $Z_{cs}$ and accessible for further experimental measurement of the Belle II and LHCb collaborations.
\end{abstract}

\date{\today}
\maketitle

\section{Introduction}
\label{sec:introduction}

The conventional quark model classified the simplest hadrons as the mesons ($q\bar{q}$) and baryons ($qqq$) and have achieved great success in the past forty years. However, in recent two decades, a large number of new hadron states beyond the conventional quark model have emerged with the development of experiments, which is a great challenge to the conventional quark model (see Refs.~\cite{Chen:2016qju,Hosaka:2016pey,Lebed:2016hpi,Esposito:2016noz,Guo:2017jvc,Ali:2017jda,Olsen:2017bmm,Karliner:2017qhf,Yuan:2018inv,Dong:2017gaw,Liu:2019zoy} for recent reviews). Among the new hadron states, the charmoniumlike states with a $c\bar{c}$ pair have been developed into a large family since the observation of $X(3872)$ in 2003~\cite{Choi:2003ue}. Different from the $X(3872)$ with neutral charge, the first charged charmoniumlike states $Z^-_c(4430)$ was observed in the $\pi^-\psi(2S)$ mass spectrum of $B\rightarrow K\pi^-\psi(2S)$ decays by the Belle collaboration in 2007 ~\cite{Belle:2007hrb, Belle:2009lvn, Belle:2013shl}, and then confirmed by the LHCb collaboration in the same process in 2014~\cite{LHCb:2014zfx}. Beside $Z^-_c(4430)$ in $\pi^-\psi(2S)$ mass spectrum, there are also other charmoniumlike states observed in other hidden charm final states, such as $Z_c^+(4050)$ and $Z_c^+(4250)$ in $\pi^+ \chi_{c1}(1P)$~\cite{Belle:2008qeq},  $Z_c^-(4240)$ in $\pi^- \psi(2S)$~\cite{LHCb:2014zfx}, $Z_c^\pm (3900)$~\cite{Ablikim:2013mio,Liu:2013dau} and $Z_c^+(4200)$~\cite{Belle:2014nuw} in $\pi^+ J/\psi$, $Z_c^- (4100)$ in $\pi^- \eta_c$~\cite{LHCb:2018oeg}, and $Z_c^\pm (4020)$ in $\pi^\pm h_c$ invariant mass spectrum~\cite{Ablikim:2013wzq}.

As the first confirmed charged charmoniumlike states, $Z_c^\pm (3900)$ was observed by the BESIII~\cite{Ablikim:2013mio} and Belle~\cite{Liu:2013dau} collaborations in the $\pi^\pm J/\psi$ invariant mass spectrum of $e^+ e^-\rightarrow \pi^+ \pi^- J/\psi$ at $\sqrt{s}=4.26$ GeV in 2013, and then confirmed by CLEO-c collaboration in the same process at $\sqrt{s}=4.17$ GeV~\cite{Xiao:2013iha}. $Z_c^\pm (3900)$ was also observed in $D^\ast \bar{D}$ invariant mass spectrum of $e^+ e^- \to \pi^\pm(D\bar{D}^\ast)^\mp$ process \cite{Ablikim:2013xfr}. As a partner of $Z_c^\pm (3900)$, $Z_c^\pm (4020)$ was discovered in the $h_c\pi^\pm$ invariant mass spectrum of $e^+ e^-\rightarrow\pi^+ \pi^- h_c$ \cite{Ablikim:2013wzq} and then observed in the $D^\ast \bar{D}^\ast$ invariant mass spectrum of $e^+ e^- \to (D^\ast \bar{D}^\ast)^\pm \pi^\mp$ process \cite{Ablikim:2013emm} by the BESIII collaboration.

It is interesting to notice that the prominent feature of one kind of charmomiumlike states, such as
$X(3872)$, $Z_c (3900)$, $Z_c (4020)$, are close to the thresholds of a pair of charmed mesons, which has attracted theorists' great interests. Since this particular property, these states have been interpreted as molecular states composed of $D^\ast \bar{D}+c.c$~\cite{Wang:2013daa,Liu:2008fh,Voloshin:1976ap,DeRujula:1976zlg,Tornqvist:1993ng,Liu:2008tn, Liu:2006df,Dong:2008gb,Wu:2021udi,Sun:2012zzd,Chen:2015igx,Xiao:2018kfx,Li:2013xia}, $D^\ast \bar{D}^\ast$~\cite{Sun:2012zzd,He:2013nwa,Chen:2013omd,Xiao:2018kfx,Li:2013xia}, respectively. By using the one-boson-exchange model (OBE), the authors considered that $X(3872)$ could be accommodated as a molecular state of $D \bar{D}^\ast+ c.c$ with isospin zero~\cite{Liu:2008tn, Liu:2008fh}, while $Z_c (3900)$ and $Z_c (4020)$ were regarded as $D\bar{D}^*+c.c$ and $D^* \bar{D}^*$  molecular states with isospin one~\cite{Sun:2012zzd,He:2013nwa}, respectively. The calculations from QCD sum rule(QCDSR) by utilizing a $D^\ast \bar{D}^\ast$ current also support the $D^\ast \bar{D}^\ast$ molecular state picture for $Z_c (4020)$~\cite{Chen:2013omd}. The productions and decays of $X(3872)$ and $Z_c (3900)$/$Z_c (4020)$~~\cite{Wu:2021udi,Chen:2015igx,Xiao:2018kfx,Li:2013xia} were investigated by using effective Lagrangian approach, the results supported the molecular state interpretation. Moreover, these charmoniumlike states were observed in the hidden charm final states. Considering the masses decay properties of these charmoniumlike state, one can find the most possible quark components of these states are $c\bar{c}q\bar{q}$, which indicates that all these charmoniumlike states could be regarded as tretraquark candidates~\cite{Maiani:2005pe,Nielsen:2006jn,Dubnicka:2010kz,Deng:2014gqa,Deng:2015lca,Brodsky:2014xia,Lebed:2017min}. Besides the QCD exotic interpretations, the charged charmoniumlike structure $Z_c^\pm (3900)$ and $Z_c^\pm (4020)$ could be reproduced through initial-single-pion-emission mechanism (ISPE)~\cite{Chen:2013coa, Chen:2013axa, Chen:2013bha, Wang:2013qwa, Chen:2012yr, Chen:2011pv}.

After the observations of $Z_c$ states, the existence of the SU(3) flavor partner of $Z_c$ states, $Z_{cs}$, has been investigated  from both theoretical and experimental sides. In Refs.~\cite{Ebert:2008kb, Ferretti:2020ewe}, some compacted tetraquark states with hidden charm and open strange were predicted, and the lowest one with $J^P=1^+$ is predicted to be around 4 GeV. Similarly, such a state was also predicted in molecular scenario~\cite{Lee:2008uy,Dias:2013qga} and ISPE mechanism~\cite{Chen:2013wca}. On the experimental side, the Belle~\cite{Belle:2007dwu,Belle:2014fgf} and BESIII~\cite{BESIII:2018iop} collaborations attempted to search $Z_{cs}$ states in the $e^+e^- \to K^+ K^- J/\psi$ process successively. Unfortunately, no obvious structures were observed in the $KJ/\psi$ invariant mass distributions due to the low statistics of the data sample.

\begin{table}[t]
 \centering
 \caption{The resonance parameters of the newly reported $Z_{cs}$ states from different collaborations ~\cite{Ablikim:2020hsk,Ablikim:2020hsk}. \label{Tab:zcs} }
 \begin{tabular}{p{2cm}<\centering p{3cm}<\centering p{2cm}}
 \toprule[1pt]
 Collaboration & Mass (MeV) & Width (MeV) \\
 \midrule[1pt]
 BES III~\cite{Ablikim:2020hsk} &         $3982.5^{+1.8}_{-2.6}\pm2.1$  & $12.8^{+5.3}_{-4.4}\pm3.0$  \\
 LHCb~\cite{Aaij:2021ivw} &  $4003 \pm 6^{+4}_{-14}$  & $131 \pm 15 \pm 26$ \\
 \bottomrule[1pt]
 \end{tabular}
 \end{table}

Recently, the experimental breakthrough of observing hidden charm and open strange states was made by the BESIII and LHCb collaborations~\cite{Ablikim:2020hsk, Aaij:2021ivw}. The BESIII collaboration reported a structure in the $K^+$ recoil-mass spectrum of process $e^+ e^-\rightarrow K^+ (D^-_s D^{\ast0}+D^{\ast-}_s D^0)$~\cite{Ablikim:2020hsk}, which is named $Z^-_{cs}(3985)$. Later, the LHCb collaboration observed $Z^+_{cs}(4000)$ in the $J/\psi K$ invariant mass spectrum of $B^+\rightarrow J/\psi\phi K^+$ process and the $J^P$ quantum numbers were determined to be $1^+$~\cite{Aaij:2021ivw}. The observed resonance parameters from different collaborations are listed in Table~\ref{Tab:zcs}. One can find the observed masses from two collaborations are very close, but the widths are very different. It should be noticed that if one consider $Z_{cs}(3985)$ as the SU(3) flavor partner of $Z_c(3900)$, there should be another states corresponding to $Z_c(4020)$, which is around 4.1 GeV. Similar to the open charm observation of $Z_c(3900)$ and $Z_c(4020)$,  only $Z_{cs}(3985)$ is expected in the process $e^+ e^-\rightarrow K^+ (D^-_s D^{\ast0}+D^{\ast-}_s D^0)$, which is consistent with the BES III measurements~\cite{Ablikim:2020hsk}. However, in the $B^+\rightarrow J/\psi\phi K^+$ process, besides $Z_{cs}(4000)$, the LHCb collaboration reported another broad structure $Z_{cs}(4220)$, which is different with the expected one near 4.1 GeV. Furthermore, it is worthwhile to mention that   the measured $J/\psi K$ invariant mass distributions near 4.1 GeV can not be well described with the broad $Z_{cs}(4000)$ and $Z_{cs}(4220)$. Further experimental fit to the LHCb data with $Z_{cs}(3985)$ and a state near 4.1 GeV may reduce the discrepancy of width from BES III and LHCb collaboration. Such possibility has also been proposed in Ref.~\cite{Yang:2020nrt}. Thus, in the present work, we assume that $Z_{cs}(4000)$ observed by LHCb collaboration should be the same states as $Z_{cs}(3985)$, and hereafter we use $Z_{cs}$ refer to this charmoniumlike state with open strange.

Similar to $Z_c (3900)/Z_c (4020)$, the observation of $Z_{cs}$ states has been stimulated theorist to put forward various pictures and explanations to probe the inner structure of $Z_{cs}$. Just as $Z_c (3900)/Z_c (4020)$ close to $D^{(*)}\bar{D}^{*}$, the observed mass of $Z_{cs}$ also locate near the threshold of $\bar{D}_s D^*/\bar{D}^*_s D$, it is natural to investigate $Z_{cs}$ in the $\bar{D}_s D^*/\bar{D}^*_s D$ molecular frame~\cite{Meng:2020ihj,Yang:2020nrt,Sun:2020hjw,Wang:2020rcx,Wang:2020htx,Dong:2020hxe,Xu:2020evn,Ozdem:2021yvo,Yan:2021tcp,Wu:2021ezz,Liu:2020nge,Chen:2020yvq}. However, the authors in Ref.~\cite{Liu:2020nge} found that $Z_{cs}$ is not a pure $\bar{D}_s D^*/\bar{D}^*_s D$ molecular state and the $\bar{D}_s D^*/\bar{D}^*_s D$ resonance assignment is also excluded in Ref.~\cite{Chen:2020yvq} by using the OBE model. Besides, $Z_{cs}$ would couple to either $\bar{D}_s D^*$ or $\bar{D}^*_s D$ and has unite charge, the quark composition is most likely $c\bar{c}s\bar{u}$. Thus, the tetraquark scenarios is also a promising  explanation, and the calculations from QCDSR~\cite{Wan:2020oxt, Wang:2020iqt, Ozdem:2021yvo} and quark model~\cite{Jin:2020yjn,Giron:2021sla,Yang:2021zhe} support $Z_{cs}$ as a compact tetraquark state. In addition to molecular and tretraquark interpretations, reflection mechanism~\cite{Wang:2020kej} and threshold effect~\cite{Ikeno:2021ptx} were proposed to reveal their exotic nature. In Refs.~\cite{Liu:2021ojf,Wu:2021ezz}, we investigated the production and hidden charm decay of $Z_{cs}$ with an effective Lagrangian approach. By studying the production of $Z_{cs}$ accompany with $P_c$ states in kaon induced reactions~\cite{Liu:2021ojf}, the cross section for $Kp\rightarrow Z_{cs}P_c$ could reach up to 10 nb. In the $\bar{D}_s D^*/\bar{D}^*_s D$ molecular frame, the dominating decay mode of $Z_{cs}$ is found to be open charm channel~\cite{Wu:2021ezz}. Under the heavy quark spin symmetry, another higher $Z^*_{cs}$ state coupling to $D^{*-}_s D^{*0}+c.c.$ should exist. The authors in Ref.~\cite{Cao:2021ton} suggest to search the $Z^*_{cs}$ in $\bar{B}^0_s\rightarrow J/\psi K^+ K^-$ at LHCb and in $e^+ e^-$ collision with the center of mass energy of 4.648 GeV.

Besides the mass and width, the LHCb collaboration also reported the fit fraction of $Z_{cs}$ in the $B^+\to J/\psi \phi K^+$ process~\cite{Aaij:2021ivw}, which is,
\begin{eqnarray}
\frac{\mathcal{B}[B^+\rightarrow Z^{+}_{cs}\phi\rightarrow J/\psi \phi K^+]}{\mathcal{B}[B^+\rightarrow J/\psi \phi K^+]}
=(9.4\pm 2.1 \pm 3.4)\%. \label{Eq:FF}
\end{eqnarray}
Considering the PDG average of the branching ratio of $B^+ \to J/\psi \phi K^+$ to be $(5.0 \pm 0.4) \times 10^{-5}$, one can conclude that the branching ratio of the cascade process is,
\begin{eqnarray}
	\mathcal{B}[B^+ \to \phi Z_{cs}^+ \to \phi (J/\psi K^+)]=(4.6\pm 2.0)\times 10^{-6}. \label{Eq:BR}
\end{eqnarray}
 How to understand such a large branching ratio of the cascade decay process is crucial to reveal the nature of $Z_{cs}$ state. Checking the decay processes of $B^+$, we notice that the branching ratio of $B^+\rightarrow D^{(*)+}_s \bar{D}^{(*)0}$ are of order of $10^{-3}$ ~\cite{ParticleDataGroup:2020ssz}.  In particular, they are $(9.0\pm0.9)\times10^{-3}$, $(7.6\pm1.6)\times10^{-3}$, $(8.2\pm1.7)\times10^{-3}$, and $(1.71\pm0.24)\%$ for $D^{+}_s \bar{D}^0$, $D^{\ast +}_s \bar{D}^0$, $D^{+}_s \bar{D}^{\ast0}$, and $D_s^{\ast+} \bar{D}^{\ast 0}$ channels, respectively ~\cite{ParticleDataGroup:2020ssz}. The charmed meson and charmed strange meson can transit into $\phi Z_{cs}^+$ by exchanging a proper charmed strange meson. Such a production mechanism will be checked in the present work. Moreover, it should be mention that, the branching ratio of $B_s^0\to D_{s}^{(\ast) +} {D}_s^{(\ast)-}$ are also sizable, the charmed strange meson pair can transit into $Z_{cs}^+ K^-$ by exchanging a proper charmed meson. Then the production of $Z_{cs}^+$ from $B_s^0$ decay will also be considered in the present work.

This work is organized as follows. After introduction, we present the model used in the present estimations of the productions  of $Z_{cs}$. The numerical results and discussions are presented in  Sec.~\ref{Sec:Num}, and Sec.~\ref{sec:summary} is devoted to a short summary.

\begin{figure}[htb]
\begin{tabular}{cc}
  \centering
 \includegraphics[width=3.5cm]{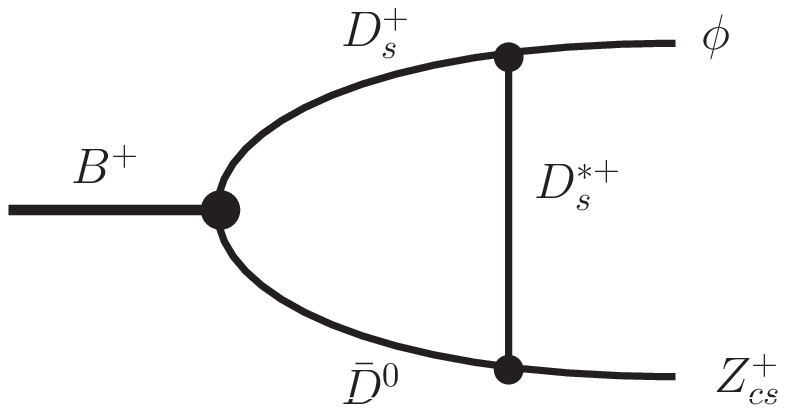}&
 \includegraphics[width=3.5cm]{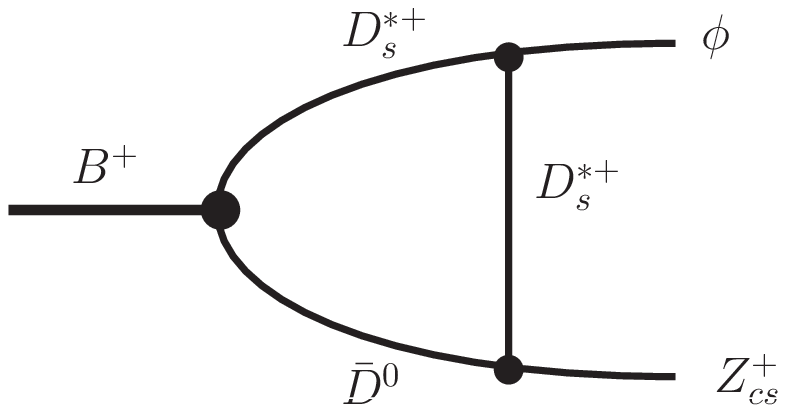}\\
 \\
 $(a)$ & $(b)$  \\
 \\
 \includegraphics[width=3.5cm]{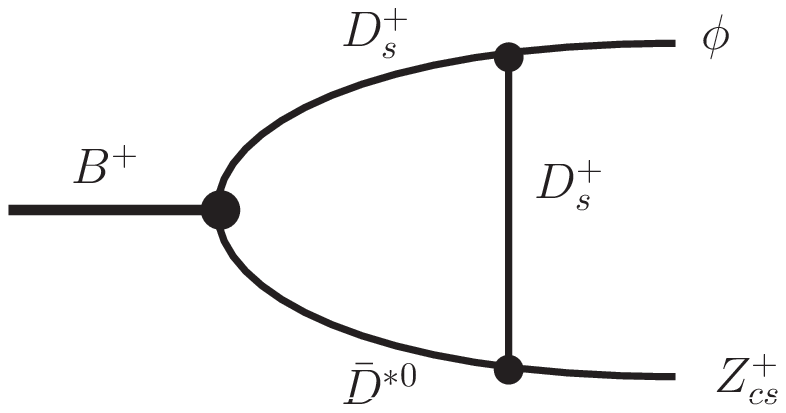}&
 \includegraphics[width=3.5cm]{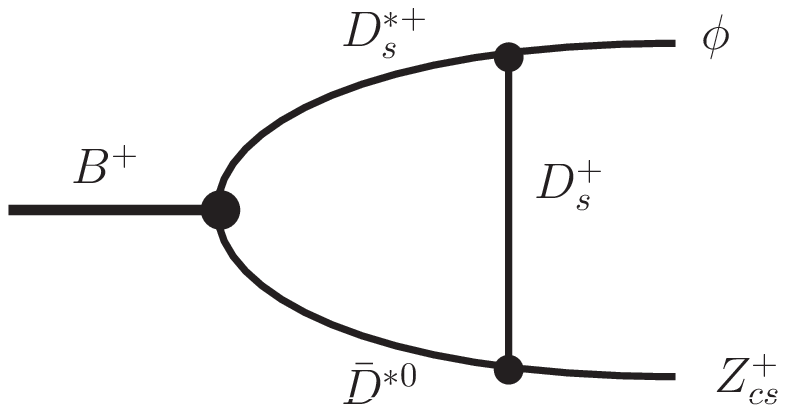}\\
 \\
  $(c)$ & $(d)$
 \end{tabular}
  \caption{Diagrams  contributing to $B^+\rightarrow \phi Z^{+}_{cs}$.}\label{Fig:Trizcsp}
\end{figure}
\begin{figure}[htb]
\begin{tabular}{cccc}
  \centering
 \includegraphics[width=2.7cm]{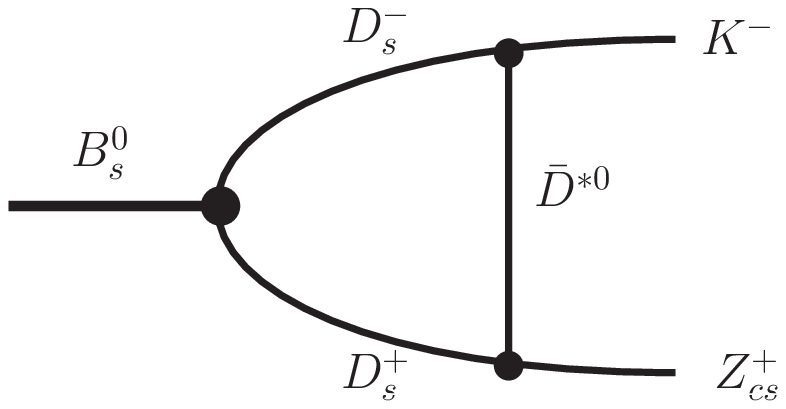}&
 \includegraphics[width=2.7cm]{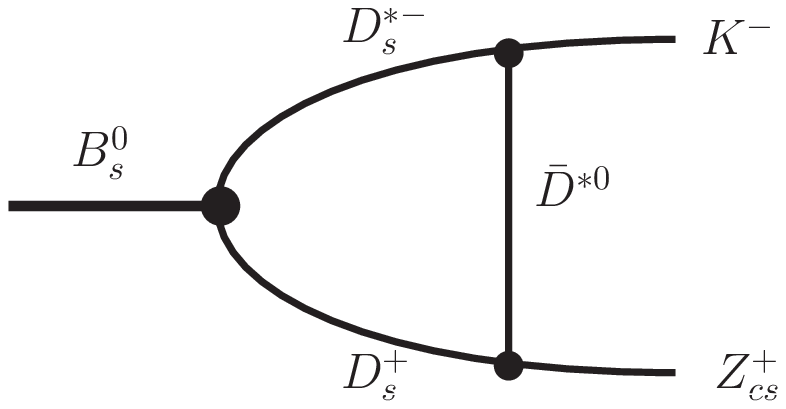}&
 \includegraphics[width=2.7cm]{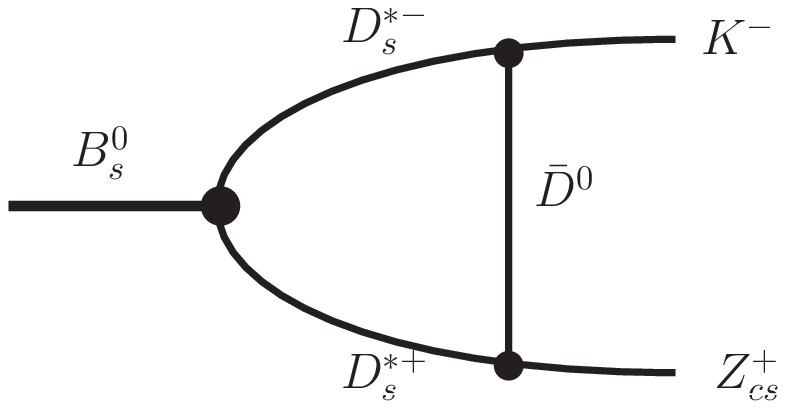}&\\
 \\
 $(e)$ & $(f)$  & $(g)$ \\
 \end{tabular}
  \caption{Diagrams  contributing to $B^0_s\rightarrow K^- Z^+_{cs}$.}\label{Fig:Trizcs}
\end{figure}

\section{Theoretical framework}
\label{sec:Sec2}

 In the $D_s^+ \bar{D}^{\ast 0} + D_s^{\ast +} \bar{D}^0 $ molecular scenario, $Z_{cs}^+$ can produced from $B^+ $ decay via the following way. The initial $B^+$ meson couples with a charmed meson and a charm-strange meson, by exchanging a proper charm-strange meson, such as $D_s^+$ or $D_s^{\ast +}$, the charmed and charm-strange mesons transit into $\phi Z_{cs}^+$ in the final state. All the possible diagrams considered in the present work are listed in Fig.~\ref{Fig:Trizcsp}. As shown in Fig.~\ref{Fig:Trizcs}, the $B_s^0$ mesons can couples to $Z_{cs}^+ K^-$ in the final states by the meson loops in a similar way.

\subsection{Effective Lagrangian}

In the present work, the diagrams in Figs. ~\ref{Fig:Trizcsp}-\ref{Fig:Trizcs} are evaluated in hadron level, where the interactions between hadrons are described by effective Lagrangians. Here, the $Z_{cs}^+$  is assumed to be a $S$-wave shallow bound states of $D_s^+ \bar{D}^{\ast 0} + D_s^{\ast +} \bar{D}^0 $ with $I(J^P)=\frac{1}{2}(1^+)$, which is,
\begin{eqnarray}
|Z^+_{cs}\rangle &=& \frac {1} {\sqrt {2}} \left(|D^{\ast+}_s \bar{D}^0\rangle+|D^+_s \bar{D}^{\ast0}\rangle \right).
\label{eq:wavef}
\end{eqnarray}
The effective coupling of $Z^+_{cs}$ to its components in terms of the following effective Lagrangian,
\begin{eqnarray}
{\cal L}_{Z_{cs}}&=&\frac {g_{Z_{cs}}} {\sqrt {2}}Z^{\dagger\mu}_{cs}\left( D^{\ast}_{s\mu} \bar{D}+ D_s \bar{D}^{\ast}_\mu \right ),
\label{eq:lagX}
\end{eqnarray}
where $g_{Z_{cs}}$ is the effective coupling constant.

As for the couplings $BD^{(*)}_s D^{(*)}$ and $B_{s}D^{(*)}_s D^{(*)}_s$, the decay amplitudes could be estimated by the naive factorization approach. The parametrized hadronic matrix elements can be obtained by applying the effective Hamiltonian at the quark level to the hadron states, which are~\cite{Cheng:2003sm,Soni:2021fky},
\begin{eqnarray}
&&\langle0|J_\mu|P(p_1)\rangle= -if_p p_{1\mu},\nonumber \\
&& \langle0|J_\mu|V(p_1,\epsilon)\rangle=f_V \epsilon_\mu m_V,\nonumber \\
&&\langle P(p_2)|J_\mu|B_{(s)}(p)\rangle  \nonumber \\
&&\quad =\Big[P_\mu-\frac{m^2_{B_{(s)}}-m^2_P}{q^2}q_\mu\Big]F_1(q^2)+\frac{m^2_{B_{(s)}}-m^2_P}{q^2}q_\mu F_0(q^2),\nonumber \\
&&\langle V(p_2,\epsilon)|J_\mu|B_{(s)}(p)\rangle
\nonumber \\
&&\quad=\frac{i\epsilon^\nu}{m_{B_{(s)}}+m_V}\Big\{i\varepsilon_{\mu\nu\alpha\beta}P^\alpha q^\beta A_V(q^2)+(m_{B_{(s)}}+m_V)^2 g_{\mu\nu}A_1(q^2)\nonumber \\
&&\quad -P_\mu P_\nu A_2(q^2)-2m_V(m_{B_{(s)}}+m_V)\frac{P_\nu q_\mu}{q^2}[A_3(q^2)-A_0(q^2)]\Big\},\nonumber
\\ \label{Eq:1}
\end{eqnarray}
where $J_\mu=\bar{q}_1 \gamma_\mu(1-\gamma_5)q_2$, $P_\mu=(p+p_2)_\mu$ and $q_\mu=(p-p_2)_\mu$. The form factor $A_3(q^2)$ is the linear combination of $A_1(q^2)$ and $A_2(q^2)$, which is~\cite{Cheng:2003sm},
\begin{eqnarray}
A_3(q^2)=\frac{m_{B_{(s)}}+m_V}{2m_V}A_1(q^2)-\frac{m_{B_{(s)}}-m_V}{2m_V}A_2(q^2).
\end{eqnarray}

With Eq.~(\ref{Eq:1}), the amplitudes of $B^0_s\rightarrow D^{(\ast)+}_s D^{(\ast)-}_s$ and $B^+ \rightarrow D^{(\ast)+}_s \bar{D}^{(\ast)0}$ are written as

\begin{eqnarray}
\mathcal{M}(B^0_s\rightarrow D^{+}_s D^{-}_s)&\equiv&\mathcal{A}^{B_s\rightarrow D_s \bar{D}_s}(p_1,p_2)\nonumber \\
\mathcal{M}(B^0_s\rightarrow D^{\ast+}_s D^{-}_s)&\equiv&\mathcal{A}^{B_s\rightarrow D^\ast_s \bar{D}_s}_\nu(p_1,p_2)\epsilon^\nu(p_1)\nonumber \\
\mathcal{M}(B^0_s\rightarrow D^{\ast+}_s D^{\ast-}_s)&\equiv&\mathcal{A}^{B_s\rightarrow D^*_s \bar{D}^*_s}_{\mu\nu}(p_1,p_2)\epsilon^\mu(p_1)\epsilon^\nu(p_2)\nonumber \\
\mathcal{M}(B^+ \rightarrow D^{+}_s \bar{D}^0)&\equiv&\mathcal{A}^{B\rightarrow D_s \bar{D}}(p_1,p_2)\nonumber \\
\mathcal{M}(B^+\rightarrow D^{\ast+}_s \bar{D}^0)&\equiv&\mathcal{A}^{B\rightarrow D^\ast_s \bar{D}}_\mu(p_1,p_2)\epsilon^\mu(p_1)\nonumber \\
\mathcal{M}(B^+\rightarrow D^{+}_s \bar{D}^{*0})&\equiv&\mathcal{A}^{B\rightarrow D_s \bar{D}^*}_\nu(p_1,p_2)\epsilon^\nu(p_2)\nonumber \\
\mathcal{M}(B^+\rightarrow D^{\ast+}_s \bar{D}^{\ast0})&\equiv&\mathcal{A}^{B\rightarrow D^*_s \bar{D}^*}_{\mu\nu}(p_1,p_2)\epsilon^\mu(p_1)\epsilon^\nu(p_2)\qquad \label{Eq:AmpWeak}
\end{eqnarray}
where the expressions of $\mathcal{A}(p_1,p_2)$, $\mathcal{A}_\nu(p_1,p_2)$ and $\mathcal{A}_{\mu\nu}(p_1,p_2)$ are collected in Appendix.~\ref{sec:appendix} for brevity.

The Lagrangians relevant to the light vector and pseudoscalar mesons can be constructed based on the heavy quark limit and chiral
symmetry~\cite{Casalbuoni:1996pg,Colangelo:2003sa,Cheng:2004ru}, which are used to describe the interactions of $D^{(*)}_s D^{(*)}K$ and $D^{(*)}_s D^{(*)}_s \phi$ in the present work, which are,
\begin{eqnarray}
 {\cal L} &=& -ig_{D^{\ast }D
{\mathcal P}}\left( D^{\dag}_i \partial_\mu {\mathcal P}_{ij} D_j^{\ast \mu}-D_i^{\ast \mu\dagger} \partial_\mu {\mathcal P}_{ij}  D_j\right) \nonumber \\
&& +\frac{1}{2}g_{D^\ast D^\ast {\mathcal P}}\varepsilon _{\mu
\nu \alpha \beta }D_i^{\ast \mu \dag}\partial^\nu {\mathcal P}_{ij}  {\overset{
\leftrightarrow }{\partial }}{\!^{\alpha }} D_j^{\ast \beta } - ig_{{D}{D}\mathcal{V}} {D}_i^\dagger {\stackrel{\leftrightarrow}{\partial}}{\!_\mu} {D}^j(\mathcal{V}^\mu)^i_j \nonumber \\
&& -2f_{{D}^*{D}\mathcal{V}} \epsilon_{\mu\nu\alpha\beta}
(\partial^\mu \mathcal{V}^\nu)^i_j
({D}_i^\dagger{\stackrel{\leftrightarrow}{\partial}}{\!^\alpha} {D}^{*\beta j}-{D}_i^{*\beta\dagger}{\stackrel{\leftrightarrow}{\partial}}{\!^\alpha} {D}^j) \nonumber
\\
&&+ ig_{{D}^*{D}^*\mathcal{V}} {D}^{*\nu\dagger}_i {\stackrel{\leftrightarrow}{\partial}}{\!_\mu} {D}^{*j}_\nu(\mathcal{V}^\mu)^i_j \nonumber \\
&& +4if_{{D}^*{D}^*\mathcal{V}} {D}^{*\dagger}_{i\mu}(\partial^\mu \mathcal{V}^\nu-\partial^\nu
\mathcal{V}^\mu)^i_j {D}^{*j}_\nu +{\rm H.c.} , \label{eq:light-meson}
 \label{eq:LDDV}
 \end{eqnarray}
 where the ${D}^{(\ast)\dagger}=(\bar{D}^{(\ast)0},D^{(\ast)-},D^{(\ast)-}_s)$ is the charmed
meson triplets, $\mathcal P$ and ${\mathcal V}_\mu$ are $3\times 3$ matrices form of pseudoscalar and vector mesons, and their concrete forms are,
\begin{eqnarray}
\mathcal{P} &=&
\left(\begin{array}{ccc}
\frac{\pi^{0}}{\sqrt 2}+\alpha\eta+\beta\eta^\prime &\pi^{+} &K^{+}\\
\pi^{-} &-\frac{\pi^{0}}{\sqrt2}+\alpha\eta+\beta\eta^\prime&K^{0}\\
K^{-} &\bar K^{0} &\gamma\eta+\delta\eta^\prime
\end{array}\right),\nonumber\\ \nonumber\\
\mathcal{V} &=& \left(\begin{array}{ccc}\frac{\rho^0} {\sqrt {2}}+\frac {\omega} {\sqrt {2}}&\rho^+ & K^{*+} \\
\rho^- & -\frac {\rho^0} {\sqrt {2}} + \frac {\omega} {\sqrt {2}} & K^{*0} \\
K^{*-}& {\bar K}^{*0} & \phi \\
\end{array}\right) ,
\end{eqnarray}
with $\alpha$ and $\beta$ are parameters related to the mixing angle $\theta$, which is $-19.1^\circ$~\cite{MARK-III:1988crp,DM2:1988bfq}.

\subsection{Decay Amplitude}

With the above effective Lagrangians, we can obtain the amplitudes for $B^+\rightarrow \phi + Z^{+}_{cs}$ corresponding to the diagrams in Fig.~\ref{Fig:Trizcsp}, which are,
\begin{eqnarray}
\mathcal{M}_{a}&=&i^3 \int\frac{d^4 q}{(2\pi)^4}\mathcal{A}^{B\rightarrow D_s \bar{D}}\Big(p_1,p_2\Big)\Big(-2f_{D_s D^*_s \phi}\varepsilon_{\mu\nu\alpha\beta}ip^\mu_3 \epsilon^\nu_{\phi}(-)\nonumber \\
&&(-i)(p_1+q)^\alpha\Big)\Big(\frac{g_{Z_{cs}}}{\sqrt{2}}\epsilon^{Z_{cs}}_\sigma\Big)\frac{-g^{\beta\sigma}+q^{\beta} q^{\sigma} /m^2_q}{q^2-m^2_q}\mathcal{F}(q^2,m_q^2),\nonumber\\
\mathcal{M}_{b}&=&i^3 \int\frac{d^4 q}{(2\pi)^4}\mathcal{A}_\mu^{B\rightarrow D^*_s \bar{D}}\Big(p_1,p_2\Big)\Big(ig_{D^\ast_s D^\ast_s \phi}g^\nu_\tau g_{\theta\nu}(-ip_{1\kappa}\nonumber\\
&&-iq_{\kappa})\epsilon^\kappa_{\phi}+4if_{D^\ast_s D^\ast_s \phi}g_{\tau\kappa}g_{\theta\nu}i(p^\kappa_3 \epsilon^\nu_{\phi}-p^\nu_3 \epsilon^\kappa_{\phi})\Big)\Big(\frac{g_{Z_{cs}}}{\sqrt{2}}\epsilon^{Z_{cs}}_\sigma\Big)\nonumber \\
&&\frac{-g^{\mu\tau}+p^{\mu}_1 p^{\tau}_1 /m^2_1}{p^2_1-m^2_1}\frac{-g^{\theta\sigma}+q^{\theta} q^{\sigma} /m^2_q}{q^2-m^2_q}\mathcal{F}(q^2,m_q^2),\nonumber\\
\mathcal{M}_{c}&=&i^3 \int\frac{d^4 q}{(2\pi)^4}\mathcal{A}_\nu^{B\rightarrow D_s \bar{D}^*}\Big(p_1,p_2\Big)\Big(-ig_{D_s D_s \phi}(-ip_{1\rho}\nonumber\\
&&-iq_\rho)\epsilon^\rho_{\phi}\Big)\Big(\frac{g_{Z_{cs}}}{\sqrt{2}}\epsilon^{Z_{cs}}_\sigma\Big)\frac{-g^{\nu\sigma}+p^{\nu}_2 p^{\sigma}_2 /m^2_2}{p^2_2-m^2_2}\mathcal{F}(q^2,m_q^2),\nonumber
\end{eqnarray}
\begin{eqnarray}
\mathcal{M}_{d}&=&i^3 \int\frac{d^4 q}{(2\pi)^4}\mathcal{A}_{\mu\nu}^{B\rightarrow D^*_s \bar{D}^*}\Big(p_1,p_2\Big)\Big(-2f_{D_s D^*_s \phi}\varepsilon_{\rho\tau\delta\xi}ip^\rho_3 \epsilon^\tau_{\phi}\nonumber \\
&&(-ip^\delta_1-iq^\delta)\Big)\Big(\frac{g_{Z_{cs}}}{\sqrt{2}}\epsilon^{Z_{cs}}_\sigma\Big)\frac{-g^{\mu\xi}+p^{\mu}_1 p^{\xi}_1 /m^2_1}{p^2_1-m^2_1}\nonumber \\
&&\frac{-g^{\nu\sigma}+p^{\nu}_2 p^{\sigma}_2 /m^2_2}{p^2_2-m^2_2}\mathcal{F}(q^2,m_q^2), \label{eq:amp2}
\end{eqnarray}
Similarly, the amplitudes for $B^0_s\rightarrow K^- + Z^+_{cs}$ corresponding to diagrams in Fig.~\ref{Fig:Trizcs} are,
\begin{eqnarray}
\mathcal{M}_{e}&=&i^3 \int\frac{d^4 q}{(2\pi)^4}\mathcal{A}^{B_s\rightarrow D_s \bar{D}_s}\Big(p_1,p_2\Big)\Big(-ig_{D_s D^\ast K}(-)ip_{3\mu}\Big)\nonumber \\
&&\Big(\frac{g_{Z_{cs}}}{\sqrt{2}}\epsilon^{Z_{cs}}_\sigma\Big)\frac{-g^{\mu\sigma}+q^{\mu} q^{\sigma} /m^2_q}{q^2-m^2_q}\mathcal{F}(q^2,m_q^2),\nonumber\\
\mathcal{M}_{f}&=&i^3 \int\frac{d^4 q}{(2\pi)^4}\mathcal{A}^{B_s\rightarrow D^*_s \bar{D}_s}_\nu\Big(p_1,p_2\Big)\Big(\frac{1}{2}g_{D^\ast_s D^\ast K}\varepsilon_{\rho\tau\kappa\xi}ip^\tau_3 (-i)\nonumber \\
&&(p_1+q)^\kappa\Big)\Big(\frac{g_{Z_{cs}}}{\sqrt{2}}\epsilon^{Z_{cs}}_\sigma\Big)\frac{-g^{\nu\xi}+p^{\nu}_1 p^{\xi}_1 /m^2_1}{p_1^2-m^2_1}\frac{-g^{\rho\sigma}+q^{\rho} q^{\sigma} /m^2_q}{q^2-m^2_q}\nonumber\\
&&\mathcal{F}(q^2,m_q^2),\nonumber\\
\mathcal{M}_{g}&=&i^3 \int\frac{d^4 q}{(2\pi)^4}\mathcal{A}^{B_s\rightarrow D^*_s \bar{D}^*_s}_{\mu\nu}\Big(p_1,p_2\Big)\Big(-ig_{D^\ast_s D K}ip_{3\rho}\Big)\Big(\frac{g_{Z_{cs}}}{\sqrt{2}}\epsilon^{Z_{cs}}_\sigma\Big)\nonumber\\
&&\frac{-g^{\mu\rho}+p^{\mu}_1 p^{\rho}_1 /m^2_1}{p_1^2-m^2_1}\frac{-g^{\nu\sigma}+p^{\nu}_2 p^{\sigma}_2 /m^2_2}{p_2^2-m^2_2}\mathcal{F}(q^2,m_q^2).\nonumber\\ \label{eq:amp1}
\end{eqnarray}

In the above amplitudes, a form factor in monopole form is adopted to represent the off-shell effect of the exchanging charmed or charm-strange mesons, and the form factor also plays the role of avoiding integration divergences, which is,
\begin{eqnarray}
\mathcal{F}(q^2,m^2) =\frac{m^2 -\Lambda^2}{q^2-\Lambda^2},\label{Eq:A1}
\end{eqnarray}
where $\Lambda=m+\alpha \Lambda_{QCD}$ with $\Lambda_{QCD}=220$ MeV. Empirically, the model parameter $\alpha$ should be of order of unity~\cite{Tornqvist:1993vu, Tornqvist:1993ng, Locher:1993cc,Li:1996yn}, but its concrete value cannot be estimated by the first principle. In practice, we usually check the rationality  of the model parameter by comparing our estimation with the corresponding experimental measurements.

\section{Numerical Results and discussions}
\label{Sec:Num}

\subsection{Coupling constants}
Considering heavy quark limit and chiral symmetry, the  coupling constants relevant to the light vector and pseudoscalar mesons are ~\cite{Casalbuoni:1996pg,Cheng:2004ru},
\begin{eqnarray}
g_{{ D}{ D}V} = g_{{ D}^*{ D}^*V}=\frac{\beta g_V}{\sqrt{2}} , \quad f_{{ D}^*{ D}V}=\frac{ f_{{ D}^*{ D}^*V}}{m_{{ D}^*}}=\frac{\lambda g_V}{\sqrt{2}} \nonumber\, , \\
g_{{D}^{*} {D} \mathcal{P}}=\frac{2 g}{f_{\pi}} \sqrt{m_{{D}} m_{{D}^{*}}}, \quad g_{{D}^{*} {D}^{*} {P}}=\frac{g_{{ D}^{*} { D} {\mathcal {P}}}}{\sqrt{m_{{D}} m_{{D}^{*}}}},
\end{eqnarray}
where the parameter $\beta=0.9$, $g_V = {m_\rho /f_\pi}$ with $f_\pi = 132$ MeV to be the decay constant of pion~\cite{Casalbuoni:1996pg}. By matching the form factor obtained from the light cone sum rule and that calculated from lattice QCD, one can obtain the parameter $\lambda = 0.56 \, {\rm GeV}^{-1} $ and $g=0.59$~\cite{Isola:2003fh}.

In general, the form factors are usually estimated in the quark model and known only in spacelike region~\cite{Cheng:2003sm}. To cover the timelike region where the physical decay processes are relevant, some method like analytically continue is needed. In Refs.~\cite{Cheng:2003sm,Soni:2021fky}, the form factors for $B_{(s)} \rightarrow D^{(*)}_{(s)}$ are parameterized as the form,xw
\begin{eqnarray}
F(Q^2)=\frac{F(0)}{1-a\zeta+b\zeta^2},\label{Eq:A1}
\end{eqnarray}
with $\zeta=Q^2/m^2_{B_s}$ and $F(0)$, $a$ and $b$ are parameters which are collected in Table~\ref{Tab:PARA1}\footnote{
In Ref.~\cite{Soni:2021fky}, the transition matrix elements of $B_s\rightarrow P/V$ are presented in a different expression, which are,
\begin{eqnarray}
\langle P(p_2)|J_\mu|B_{s}(p)\rangle&=&F_+(q^2)P^\mu+F_-(q^2)q^\mu,\nonumber\\
\langle V(p_2,\epsilon)|J_\mu|B_{(s)}(p)\rangle&=&\frac{\epsilon_\nu}{m+m_2}[-g^{\mu\nu}P\cdot qA_0(q^2)+P^\mu P^\nu A_+(q^2)\nonumber\\
&&+q^\mu P^\nu(q^2)+i\varepsilon^{\mu\nu\alpha\beta}P_\alpha q_\beta V(q^2)].
\nonumber
\end{eqnarray}
which are identical with the expression in Eq.~(\ref{Eq:1}). By comparing the above paramaterizaiton with the one in Eq.~(\ref{Eq:1}), one can find the form factors has the following relation:
\begin{eqnarray}
F_1(q^2)&=&F_+(q^2),\nonumber\\
F_0(q^2)&=&\frac{q^2}{m^2_{B_s}-m^2_P}F_-(q^2)+F_+(q^2),\nonumber\\
A_V(q^2)&=&-iV(q^2),\ A_2(q^2)=iA_+(q^2),  \nonumber\\
A_1(q^2)&=&\frac{iP\cdot q}{(m_{B_s}+m_V)^2}A_0(q^2),\nonumber\\
A_3(q^2)-A_0(q^2)&=&\frac{iq^2}{2m_V(m_{B_s}+m_V)}A_-(q^2).
\end{eqnarray}
}.

\begin{table}[t]
\begin{center}
\caption{The values of the parameters $F(0)$, $a$ and $b$ in the form factors of $B\rightarrow D^{(*)}$~\cite{Cheng:2003sm} and $B_s\rightarrow D^{(*)}_s$~\cite{Soni:2021fky}.
}\label{Tab:PARA1}
  \setlength{\tabcolsep}{1.8mm}{
\begin{tabular}{cccccccc}
  \toprule[1pt]
  $F$ & $F(0)$ & $a$ & $b$ & $F$ & $F(0)$ & $a$ & $b$ \\
  \midrule[1pt]
  $F_0$ & 0.67 & 0.65 & 0.00 & $F_1$ & 0.67 & 1.25 & 0.39 \\
  $A_V$ & 0.75 & 1.29 & 0.45 & $A_0$ & 0.64 & 1.30 & 0.31 \\
  $A_1$ & 0.63 & 0.65 & 0.02 & $A_2$ & 0.61 & 1.14 & 0.52 \\
  \midrule[1pt]
  $F_+$ & 0.770 & 0.837 & 0.077 & $F_-$ & -0.355 & 0.855 & 0.083 \\
  $A_+$ & 0.630 & 0.972 & -0.092 & $A_-$ & -0.756 &  1.001 &  0.116 \\
  $A_0$ & 1.564 & 0.442 & -0.178 & $V$ & 0.743 & 1.010 & 0.118 \\
  \bottomrule[1pt]
\end{tabular}}
\end{center}
\end{table}
\begin{table}[htb]
\begin{center}
\caption{Values of the parameters $\Lambda_1$ and $\Lambda_2$ obtained by fitting the form factor. }\label{Table:PARA2}
  \setlength{\tabcolsep}{0.5mm}{
\begin{tabular}{ccccccccccc}
 \toprule[1pt] 
  Process& Parameter&   $A_V$ & $A_0$& $A_1$ & $A_2$ &   $F_0$ & $F_1$ &\\
  \midrule[1pt]
 \multirow{2}{*}{$B\rightarrow D^{(*)}$}                 &$\Lambda_1$ & 6.32 & 5.32 & 7.83   & 7.35 &   7.75 & 6.53& \\
                &$\Lambda_2$ & 7.00& 9.41  & 10.99 & 7.35 &  11.00 &  6.84 &    \\
  \midrule[1pt]
  Process& Parameter&   $A_+$ & $A_-$ & $A_0$ & $V$&  $F_+$ & $F_-$ &\\
  \midrule[1pt]
 \multirow{2}{*}{$B_s\rightarrow D_s^{(*)}$}    &$\Lambda_1$ & 5.48   & 5.77 & 9.75   & 5.74 &   6.30 & 6.25& \\
                  &$\Lambda_2$ & 18.00  & 14.63  & 11.00 & 14.61 &  15.92 &  15.58&    \\
\bottomrule[1pt]
\end{tabular}}
\end{center}
\end{table}

In order to avoid ultraviolet divergence in the loop integrals and evaluate the loop integrals with Feynman parameterization methods, we further parameterize the form factors in the form
\begin{eqnarray}
F(Q^2)=F(0)\frac{\Lambda^2_1}{Q^2-\Lambda^2_1}\frac{\Lambda^2_2}{Q^2-\Lambda^2_2}.\label{Eq:A2}
\end{eqnarray}
where the values of $\Lambda_1$ and $\Lambda_2$ are obtained by fitting Eq.~(\ref{Eq:A1}) with Eq.~(\ref{Eq:A2}) and the fitted parameter values are list in Table~\ref{Table:PARA2}.

In Ref.~\cite{Wu:2021ezz}, we investigated the  decay properties of $Z^+_{cs}$ via triangle loop mechanism by using an effective Lagrangian approach, the coupling constant $g_{Z_{cs} D_s^\ast D}$ were determined to be $6.0\sim6.7$, which is weakly dependent on the model parameter. In the following, we take $g_{Z_{cs} D_s^\ast D}=6$ to roughly estimate the branching ratios of $B^+\rightarrow \phi +Z^{+}_{cs}$ and $B^0_s\rightarrow K^- + Z^+_{cs}$.

\begin{figure}[b]
  \centering
 \includegraphics[width=8.5cm]{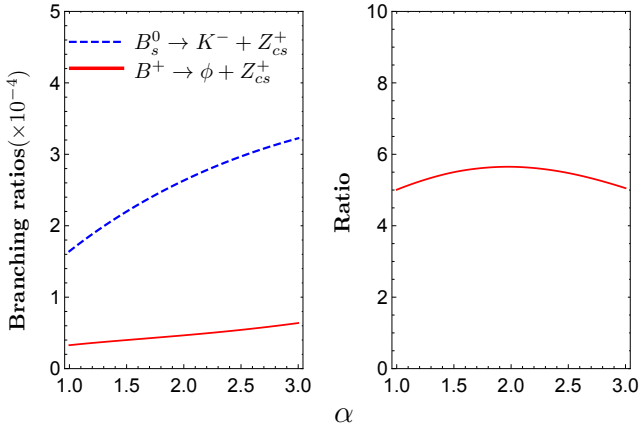}
  \caption{Branching ratios of $B^+\rightarrow \phi Z^{+}_{cs}$ and $B^0_s\rightarrow K^-  Z^+_{cs}$ in unit of $10^{-4}$ (left panel) and their ratio (right panel) depending on parameter $\alpha$.}\label{Fig:Zcsbr}
\end{figure}
\subsection{Branching ratios}

Our estimations of the branching ratio of $B^+ \to \phi Z_{cs}^+$ depending on the model parameter $\alpha$ are present in the left panel of  Fig.~\ref{Fig:Zcsbr}. As shown in the figure,  the branching ratio of $B^+ \to \phi Z_{cs}^+$ is $(4.66^{+1.72}_{-1.38})\times 10^{-5}$, where the center value is estimated by taking $\alpha=2$ and the uncertainties are resulted from the variation of model parameter. In Ref.~\cite{Wu:2021ezz}, our estimation indicated that the branching ratio of $Z_{cs}^+ \to J/\psi K^+$ was $(4.0^{+4.3}_{-2.7})\%$. Considering $Z_{cs}^+$ to be a narrow resonance, one can roughly estimate the branching ratio of the cascade process by the product of the branching ratio of $B^+ \to \phi Z_{cs}^+$ and $Z_{cs}^+ \to J/\psi K^+$, which is,
\begin{eqnarray}
&&\mathcal{B}[B^+ \to \phi Z_{cs}^+ \to \phi K^+ J/\psi] =(1.86^{+2.12}_{-1.37})\times 10^{-6},
\end{eqnarray}
which is comparable to the experimental measurement from the LHCb collaboration as shown in Eq.~(\ref{Eq:BR})~\cite{Aaij:2021ivw}. As mentioned before, experimental measured branching ratio is obtained by fitting the experimental data with a broad $Z_{cs}$ states. A reanalyze to the data with a narrow $Z_{cs}$ and an additional new states near 4.1 GeV is expected.

Besides the observed channel, we also propose to search $Z_{cs}$ state in the $B_s^0 \to J/\psi K^+ K^-$ proces. In the observed $B^+\rightarrow J/\psi \phi K^+$ process, the resonance contributions could come from $J/\psi\phi$, $J/\psi K$, and $\phi K$ invariant mass spectrum. However, in the $B_s^0 \to J/\psi K^+ K^-$ process, the dominant contributions should comes from the $KK$ resonance and $J/\psi K$ resonance, i.e., the $Z_{cs}$ states. Thus, on the experimental side, the $B_s^0 \to J/\psi K^+ K^-$ may be a more cleaner process of searching $Z_{cs}$ states. Our estimation of the branching ratio of $B^0_s\rightarrow K^- Z^+_{cs}$ depending on the model parameter is also presented in Fig.~\ref{Fig:Zcsbr}. The branching ratio is estimated to be $(2.63^{+0.59}_{-0.99})\times10^{-4}$. In the figure, one can find the parameter dependences of the branching ratios are very similar, thus, we can estimate the ratio of the branching fractions of these two process as,
\begin{eqnarray}
	R_{B_s/B}= \frac{\mathcal{B}[B_s^0 \to K^- Z_{cs}^+]}{\mathcal{B}[B^+ \to \phi Z_{cs}^+]}=2.9\sim8.1,
\end{eqnarray}
which weakly depend on the model parameter. Moreover, the PDG average of the branching ratio $B_s^0 \to J/\psi K^+ K^-$ is $(7.9\pm0.7)\times 10^{-4}$~\cite{ParticleDataGroup:2020ssz}. With the branching ratio of $B_s^0 \rightarrow K^- Z_{cs}^+$ in the present work and the one of $Z_{cs} \to J/\psi K$ in Ref.~\cite{Wu:2019vbk}, the production ratio of $Z_{cs}$ in $B^0_s\rightarrow J/\psi K^+ K^-$ is
\begin{eqnarray}
&&\frac{\mathcal{B}[B^0_s\rightarrow Z^-_{cs}K^+ +c.c.\rightarrow J/\psi K^+ K^-]}{\mathcal{B}[B^0_s\rightarrow J/\psi K^+ K^-]}\nonumber\\
&\simeq&\frac{\mathcal{B}[B^0_s\rightarrow Z^-_{cs}K^+ +c.c.]\times\mathcal{B}[Z^-_{cs}\rightarrow J/\psi K^-]}{\mathcal{B}[B^0_s\rightarrow J/\psi K^+ K^-]}\nonumber\\
&=&(2.66^{+2.93}_{-2.07})\%,
\end{eqnarray}
where the interference between $Z_{cs}^+$ and $Z_{cs}^-$ is neglected. The above production ratio could be tested by future experiment.

\section{Summary}
\label{sec:summary}
Recently, the BESIII and LHCb collaboration reported the observation of a hidden charm and open strange states, which is $Z_{cs}$. The mass of this newly observed charmoniumlike state is close to the threshold of $D_s^\ast D$, which could be a good candidates of molecular state composed of $D_s \bar{D}^\ast+D_s^\ast \bar{D}$. In the molecular scenario, the mass spectrum and decay properties have been investigated extensively. Besides the resonance parameters of $Z_{cs}$, the LHCb collaboration has also reported the branching ratio of the cascade decay process $B^+ \to \phi Z_{cs}^+ \to \phi K^+ J/\psi$, which is an order of $10^{-6}$. How to understand the production properties of $Z_{cs}$ is important to reveal its inner structure. In the present work, we estimate the production of $Z_{cs}^+$ from $B^+$ decay by considering the triangle meson loop contributions. Our estimations indicate that the branching ratio of $B^+ \to \phi Z_{cs}^+$ is of order of $10^{-5}$. Together with the decay properties $Z_{cs}$ in our previous work, we find the estimated branching ratio of $B^+ \to \phi Z_{cs}^+ \to \phi K^+ J/\psi$ is of order $10^{-7} \sim 10^{-6}$, which is comparable with the measurement from LHCb collaboration.

Besides the $B^+ \to \phi Z^+_{cs}$ process, our estimations indicate the branching ratio of $B_{s}^0 \to K^- Z_{cs}^+$ is about 5 times of the one of $B^+ \to \phi Z_{cs}^+$, which indicate that the $B_s^0 \to K^\pm  Z^\mp_{cs} \to K^+ K^- J/\psi$ may be a better process of searching $Z_{cs}$ and accessible for the experimental measurement of the Belle II and LHCb collaborations.

\section*{Acknowledgement}
Q. W is grateful to Professor Shi-Lin Zhu for very helpful discussions. This work is supported by the National Natural Science Foundation of China (NSFC) under Grant No.11775050, 12175037, 11835015, and 12075133. It is also partly supported by Taishan Scholar Project of Shandong Province (Grant No. tsqn202103062), the Higher Educational Youth Innovation Science and Technology Program Shandong Province (Grant No. 2020KJJ004)
\\
\\
\begin{appendix}
\section{The expressions of $\mathcal{A}(p_1,p_2)$, $\mathcal{A}_\nu(p_1,p_2)$ and $\mathcal{A}_{\mu\nu}(p_1,p_2)$}\label{sec:appendix}
Here we collect all the function used in Eq. ~(\ref{Eq:AmpWeak}), which are,
\begin{eqnarray}
\mathcal{A}^{B_s\rightarrow D_s \bar{D}_s}(p_1,p_2)&=&-\frac{iG_F}{\sqrt{2}}V_{cb}V^\ast_{cs}a_1 f_{D_s}(m^2_{B_s}-m^2_{D_s})F_0^{B_s D_s}(p^2_1)\nonumber \\
\mathcal{A}^{B_s\rightarrow D^*_s \bar{D}_s}_\nu(p_1,p_2)&=&\frac{G_F}{\sqrt{2}}V_{cb}V^\ast_{cs}a_1 f_{D_s}\frac{1}{m_{B_s}+m_{D^\ast_s}}\nonumber \\
&&\times\Big\{(m_{B_s}+m_{D^\ast_s})^2 g_{\mu\nu}p^\mu_{2}A_1^{B_s D^\ast_s}(p^2_2)\nonumber \\
&&-(2p_1+p_2)_\mu (2p_1+p_2)_\nu p^\mu_{2} A_2^{B_s D^\ast_s}(p^2_2)\nonumber \\
&&-2m_{D^\ast_s}(m_{B_s}+m_{D^\ast_s})(2p_1+p_2)_\nu\nonumber \\
&&\times[A_3^{B_s D^\ast_s}(p^2_2)-A_0^{B_s D^\ast_s}(p^2_2)]\Big\}\nonumber \\
\mathcal{A}^{B_s\rightarrow D^*_s \bar{D}^*_s}_{\mu\nu}(p_1,p_2)&=&\frac{G_F}{\sqrt{2}}V_{cb}V^\ast_{cs}a_1 f_{D^*_s}m_{D^\ast_s}\frac{i}{m_{B_s}+m_{D^\ast_s}}\nonumber \\
&&\times\Big\{i\varepsilon_{\mu\nu\alpha\beta}(p_1+2p_2)^\alpha p^\beta_1 A_V^{B_s D^\ast_s}(p^2_1)\nonumber \\
&&+(m_{B_s}+m_{D^\ast_s})^2 g_{\mu\nu}A_1^{B_s D^\ast_s}(p^2_1)\nonumber \\
&&-(p_1+2p_2)_\mu (p_1+2p_2)_\nu A_2^{B_s D^\ast_s}(p^2_1)\Big\}\nonumber \\
\mathcal{A}^{B\rightarrow D_s \bar{D}}(p_1,p_2)&=&-\frac{iG_F}{\sqrt{2}}V_{cb}V^\ast_{cs}a_1 f_{D_s}(m^2_{B}-m^2_{D})F_0^{B D}(p^2_1)\nonumber \\
\mathcal{A}_\mu^{B\rightarrow D^*_s \bar{D}}(p_1,p_2)&=&\frac{2G_F}{\sqrt{2}}V_{cb}V^\ast_{cs}a_1 f_{D^*_s}m_{D^\ast_s} p_{2\mu} F_1^{B D}(p^2_1)\nonumber \\
\mathcal{A}^{B\rightarrow D_s \bar{D}^*}_\nu(p_1,p_2)&=&\frac{G_F}{\sqrt{2}}V_{cb}V^\ast_{cs}a_1 f_{D_s}\frac{1}{m_{B}+m_{D^\ast}}\nonumber \\
&&\times\Big\{(m_{B}+m_{D^\ast})^2 g_{\mu\nu}p^\mu_{1}A_1^{B D^\ast}(p^2_1)\nonumber \\
&&-(p_1+2p_2)_\mu (p_1+2p_2)_\nu p^\mu_{1} A_2^{B D^\ast}(p^2_1)\nonumber \\
&&-2m_{D^\ast}(m_{B}+m_{D^\ast})(p_1+2p_2)_\nu\nonumber \\
&&\times[A_3^{B D^\ast}(p^2_1)-A_0^{B D^\ast}(p^2_1)]\Big\}\nonumber \\
\mathcal{A}^{B\rightarrow D^*_s \bar{D}^*}_{\mu\nu}(p_1,p_2)&=&i\frac{G_F}{\sqrt{2}}V_{cb}V^\ast_{cs}a_1 f_{D^*_s}\frac{m_{D^\ast_s}}{m_{B_s}+m_{D^\ast}}\nonumber \\
&&\times\Big\{i\varepsilon_{\mu\nu\alpha\beta}(p_1+2p_2)^\alpha p^\beta_1 A_V^{B D^\ast}(p^2_1)\nonumber \\
&&+(m_{B}+m_{D^\ast})^2 g_{\mu\nu}A_1^{B D^\ast}(p^2_1)\nonumber \\
&&-(p_1+2p_2)_\mu (p_1+2p_2)_\nu A_2^{B D^\ast}(p^2_1)\Big\}\nonumber \\
\end{eqnarray}
\end{appendix}


\end{document}